\documentclass[final,3p,times,twocolumn]{elsarticle}
 \biboptions{comma,sort&compress}
\usepackage{here}
 \usepackage{graphicx}
  \usepackage{epsfig}
\usepackage{amsmath}
\usepackage{amssymb}
\usepackage{float}
\usepackage[rflt]{floatflt}

\def\beq{\begin{equation}}
\def\eeq{\end{equation}}
\def\bea{\begin{eqnarray}}
\def\eea{\end{eqnarray}}

\journal{Nucl. Phys. B (Proc. Suppl.)}

\begin{document}

\begin{frontmatter}

%% Title, authors and addresses

%% use the tnoteref command within \title for footnotes;
%% use the tnotetext command for the associated footnote;
%% use the fnref command within \author or \address for footnotes;
%% use the fntext command for the associated footnote;
%% use the corref command within \author for corresponding author footnotes;
%% use the cortext command for the associated footnote;
%% use the ead command for the email address,
%% and the form \ead[url] for the home page:
%%
%% \title{Title\tnoteref{label1}}
%% \tnotetext[label1]{}
%% \author{Name\corref{cor1}\fnref{label2}}
%% \ead{email address}
%% \ead[url]{home page}
%% \fntext[label2]{}
%% \cortext[cor1]{}
%% \address{Address\fnref{label3}}
%% \fntext[label3]{}

\title{{\footnotesize DESY 14--153, DO--TH 14/21, SFB/CPP--14--68 , LPN 14--111}\\
3-loop heavy flavor Wilson coefficients in deep-inelastic scattering\tnoteref{label1}}

\tnotetext[label1]{
This work was supported in part by DFG Sonderforschungsbereich Transregio 9, Computergest\"utzte Theoretische Teilchenphysik,
Studienstiftung
des Deutschen Volkes, the Austrian Science Fund (FWF) grants P20347-N18 and SFB F50 (F5009-N15), the European
Commission through contract PITN-GA-2010-264564 ({LHCPhenoNet}) and PITN-GA-2012-316704 ({HIGGSTOOLS}), by the
Research Center ``Elementary Forces and Mathematical Foundations (EMG)'' of J.~Gutenberg University Mainz and
DFG, and by FP7 ERC Starting Grant  257638 PAGAP.}

%% use optional labels to link authors explicitly to addresses:
\author[risc]{J. Ablinger}
\address[risc]{RISC, Johannes Kepler Universit\"at, Linz, Austria.}

\author[desy]{A. Behring}
\address[desy]{Deutsches Elektronen-Synchrotron DESY, Zeuthen, Germany}

\author[desy]{J. Bl\"umlein}

\author[desy]{A. De Freitas\corref{speaker}}
\cortext[speaker]{Speaker}

\author[risc]{H. Hasselhuhn}

\author[prisma]{A. von Manteuffel}
\address[prisma]{PRISMA Cluster of Excellence, 
Institut f\"ur Physik, Johannes Gutenberg-Universit\"at, Mainz, Germany} 

\author[desy]{C. Raab}

\author[desy]{M. Round}

\author[risc]{C. Schneider}

\author[risc,desy,ihes]{F. Wi\ss{}brock}
\address[ihes]{Institut des Hautes \'Etudes Scientifiques, IHES, Bures-sur-Yvette, France}

\begin{abstract}
%% Text of abstract
\noindent
We present our most recent results on the calculation of the heavy flavor contributions to deep-inelastic scattering at 3-loop order
in the large $Q^2$ limit, where the heavy flavor Wilson coefficients are known to factorize into light flavor Wilson coefficients
and massive operator matrix elements. We describe the different techniques employed for the calculation and show the results in the case 
of the heavy flavor non-singlet and pure singlet contributions to the structure function $F_2(x,Q^2)$.
\end{abstract}

\begin{keyword}
Deep-inelastic scattering, heavy flavor
%% keywords here, in the form: keyword \sep keyword

%% MSC codes here, in the form: \MSC code \sep code
%% or \MSC[2008] code \sep code (2000 is the default)

\end{keyword}

\end{frontmatter}

%%
%% Start line numbering here if you want
%%
% \linenumbers

%% main text
%---------------------------------------------------------------------------------------------------------------------------------------
\section{Introduction}
\label{intro}
%---------------------------------------------------------------------------------------------------------------------------------------

\noindent
The deep-inelastic precision data from HERA allow determinations of $\alpha_s$ at the 1\% level \cite{Bethke:2011tr}, and 
precise measurements of the mass of the charm quark \cite{Alekhin:2012vu} and parton distribution functions. This applies to an even
further extent to the proposed deep-inelastic scattering (DIS) experiments to be carried out in the future at new facilities, such as the 
EIC \cite{EIC} and the LHeC \cite{LHeC}, reaching much higher luminosities or energies than those available at HERA \cite{hera}. 
The corresponding analyses demand the knowledge of the anomalous dimensions and Wilson coefficients at 3-loop order for the structure 
function 
$F_2(x,Q^2)$, including the heavy flavor corrections. For $Q^2 \gg m^2$, where $m$ is the mass of the heavy
quark, the massive Wilson coefficients are known to factorize into the massless ones and massive operator matrix elements, 
allowing the computation of $F_2(x,Q^2)$ to the 1\% level \cite{Buza:1995ie}. The 3-loop light flavor Wilson coefficients are known 
\cite{MVV2005}. Thus, it remains to calculate all the contributing massive OMEs. 

In these proceedings we discuss the
progress we have made in recent times in the calculation of these quantities 
\cite{OurPapers1,Behring:2014eya,TF2paper,hyperlogs,nonsinglet,PSpaper}, 
with special emphasis on the non-singlet \cite{nonsinglet} and pure singlet contributions \cite{PSpaper}. By now, six out of eight 
operator matrix elements have been computed by us, namely,
$A_{qq,Q}^{(3), \rm PS}$, $A_{qg,Q}^{(3)}$, $A_{gq}^{(3)}$, $A_{qq,Q}^{(3), \rm NS}$, $A_{qq,Q}^{(3), \rm NS, \,  TR}$ and $A_{Qq}^{(3), 
\rm PS}$.\footnote{For the notation see Ref.~\cite{Bierenbaum:2009mv}.} Only the operator matrix elements $A_{gg,Q}^{(3)}$ and
$A_{Qg}^{(3)}$ remain to be calculated, although in these cases, we also have partial results for some of the color factors \cite{TF2paper}. 
In the region $Q^2 \gg m^2$ also all Wilson coefficients for the structure function $F_L(x,Q^2)$ were calculated 
\cite{Behring:2014eya,Blumlein:2006mh}.
Using these operator matrix elements, we have computed
the massive Wilson coefficients $L_{q,(2,L)}^{\rm NS}$, $L_{q,(2,L)}^{\rm PS}$, $L_{g,(2,L)}^{\rm S}$ and $H_{q,(2,L)}^{\rm PS}$ at NNLO. Once $A_{Qg}^{(3)}$ is available, we will be able 
to obtain also $H_{g,(2)}^{\rm S}$.

With the calculation of these operator matrix elements, we also obtain as a by-product the terms proportional to $T_F$ of the 3-loop anomalous 
dimensions. In the case of transversity, we performed the first calculation ab initio. Our 
results agree with those given in the 
literature.

The operator matrix elements also yield the heavy-to-light transition relations in the variable flavor number scheme to 3-loop order. 
These relations define the parton distributions for $(n_f+1)$ flavors from those of $n_f$ light flavors by the OMEs.
In particular, we now have the relation corresponding to $f_k(n_f+1, \mu^2) + f_{\overline{k}}(n_f+1, \mu^2)$ in the flavor non-singlet 
case.

In the next Section, we describe the different methods we used for the calculation of the operator matrix elements and the required Feynman integrals. 
In Section \ref{results}, we discuss our results and show the non-singlet and pure singlet contributions to $F_2(x,Q^2)$. 
Finally, in Section \ref{conclusions} we give some conclusions.

%---------------------------------------------------------------------------------------------------------------------------------------
\section{Calculation of the operator matrix elements}
\label{OMEcalc}
%---------------------------------------------------------------------------------------------------------------------------------------

\noindent
The operator matrix elements are computed in terms of Feynman diagrams using the standard Feynman rules of QCD, together with those for 
local operator insertions \cite{Bierenbaum:2009mv}.
We generated the diagrams using {\tt QGRAF} \cite{Nogueira:1991ex}, the output of which was processed using a {\tt FORM} \cite{Tentyukov:2007mu} program \cite{Bierenbaum:2009mv}, 
which ultimately allowed us to express the diagrams as a linear combination of scalar integrals.

The number of scalar integrals required for the calculation of the OMEs is quite large. We used integration by parts identities in order to express all integrals
in terms of a relatively small set of master integrals. For this purpose we used the {\tt C++} program {\tt Reduze2} \cite{reduze}, which 
implements Laporta's algorithm \cite{laporta}.
Since this algorithm requires the integrals to be identified by definite indices, we rewrite the operator insertions by introducing an auxiliary variable
$x$, multiplying the integrals by $x^N$ and summing in $N$. For example, a line insertion is rewritten as \cite{Ablinger:2012qm}
\begin{equation}
(\Delta \cdot k)^{N-1} \rightarrow \sum^{\infty}_{N=1} x^{N-1} (\Delta \cdot k)^{N-1} = \frac{1}{1-x \Delta \cdot k} \, ,
\label{N2x1}
\end{equation}
where $k$ is the momentum going through the line, and $\Delta$ is a light-like vector. In this way, the operator insertion is turned into an
artificial propagator, making the application of Laporta's algorithm possible. Similarly, the 3-point and 4-point vertex insertions
can be re-expressed in terms of products of such artificial propagators.

The master integrals were calculated using a variety of methods, namely,

\begin{itemize}
\item Hypergeometric functions \cite{SLATER}
\item Mellin-Barnes integral representations \cite{MELB}.
\item Hyperlogarithms for convergent integrals, \cite{Brown:2008um,hyperlogs} 
\item Differential equations \cite{DEQ}.
\end{itemize}

\noindent
The representations lead to multiple nested sums which are summed using algorithms based on difference fields \cite{Karr:81}, implemented 
in the packages {\tt Sigma} \cite{SIG1}, {\tt HarmonicSums} \cite{harmonicsums,Ablinger:2013cf}, {\tt EvaluateMultiSums} and {\tt 
SumProduction} \cite{EMSSP}.
The choice of method depends on the complexity of the integral under consideration. The simplest integrals are calculated using
hypergeometric functions, while the most complicated integrals we have encountered so far have been computed in a systematic manner
using the differential equation method. This method ultimately leads to difference equations that are then solved using the packages
{\tt Sigma} and {\tt OreSys} \cite{ORESYS}.

%---------------------------------------------------------------------------------------------------------------------------------------
\section{Results}
\label{results}
%---------------------------------------------------------------------------------------------------------------------------------------

\noindent
All OMEs calculated so far have been found to be expressible in terms of (generalized) nested harmonic sums. In particular,
$A_{qq,Q}^{(3), \rm PS}$, $A_{qg,Q}^{(3)}$, $A_{gq}^{(3)}$, $A_{qq,Q}^{(3), \rm NS}$ and $A_{qq,Q}^{(3), \rm NS, \,  TR}$ can all be 
expressed solely in terms of
standard harmonic sums defined by \cite{HSUM}
%---------------------------------------------------------------------------------------------------------------------------------------
\begin{eqnarray}
S_{b,\vec{a}}(N) = \sum_{k=1}^N \frac{({\rm sign}(b))^k}{k^{|b|}} S_{\vec{a}}(k),
\label{Hsums}
\end{eqnarray}
%---------------------------------------------------------------------------------------------------------------------------------------
with $S_\emptyset = 1$, $b, a_i \in \mathbb{Z} \backslash\{0\}$.
In the case of $A_{Qq}^{(3), \rm PS}$, we find for the first time in DIS generalized harmonic sums in the final result. 
They are defined by \cite{Moch:2001zr,Ablinger:2013cf}
%---------------------------------------------------------------------------------------------------------------------------------------
\begin{eqnarray}
S_{b,\vec{a}}\big(c, \vec{d} \hspace*{0.7mm}\big)(N) = \sum_{k=1}^N \frac{c^k}{k^b} S_{\vec{a}}\big(\vec{d} 
\hspace*{0.7mm}\big)(k),~~
\label{GHsums}
\end{eqnarray}
%---------------------------------------------------------------------------------------------------------------------------------------
where $S_\emptyset = 1$, $b, a_i \in \mathbb{N} \backslash \{0\}$, $c, d_i \in \mathbb{Z}  \backslash \{0\}$. We have also found the emergence of inverse binomial
sums \cite{hyperlogs,Ablinger:2014bra}, such as
%---------------------------------------------------------------------------------------------------------------------------------------
\begin{equation}
\binom{2 N}{N} \sum_{k=0}^N \frac{k^4 2^k}{\binom{2 k}{k}} S_1(k),
\label{Bsums1}
\end{equation}
%---------------------------------------------------------------------------------------------------------------------------------------
or
%---------------------------------------------------------------------------------------------------------------------------------------
\begin{equation}
\sum_{i=1}^N \binom{2 i}{i} (-2)^i \sum_{j=1}^i \frac{1}{j \binom{2 j}{j}} S_{1,2}\left(\frac{1}{2},-1;j\right).
\label{Bsums2}
\end{equation}
%---------------------------------------------------------------------------------------------------------------------------------------
This type of sums appears, e.g., in the terms proportional to $T_F^2$ in $A_{gg}^{(3)}$, as shown in \cite{TF2paper}.

The Mellin inversion of these
quantities leads to harmonic polylogarithms in the case of Eq. (\ref{Hsums}). In the case of Eq. (\ref{GHsums}), they also lead to 
generalized harmonic polylogarithms \cite{Ablinger:2013cf} that we were able to re-express in terms of standard harmonic polylogarithms in 
a new variable $y=1-2 x$.
On the other hand, the Mellin inversion of inverse binomial sums leads to iterated integrals over square root-valued alphabets.
%---------------------------------------------------------------------------------------------------------------------------------------
\begin{figure}[H] 
\centerline{\includegraphics[width=7cm]{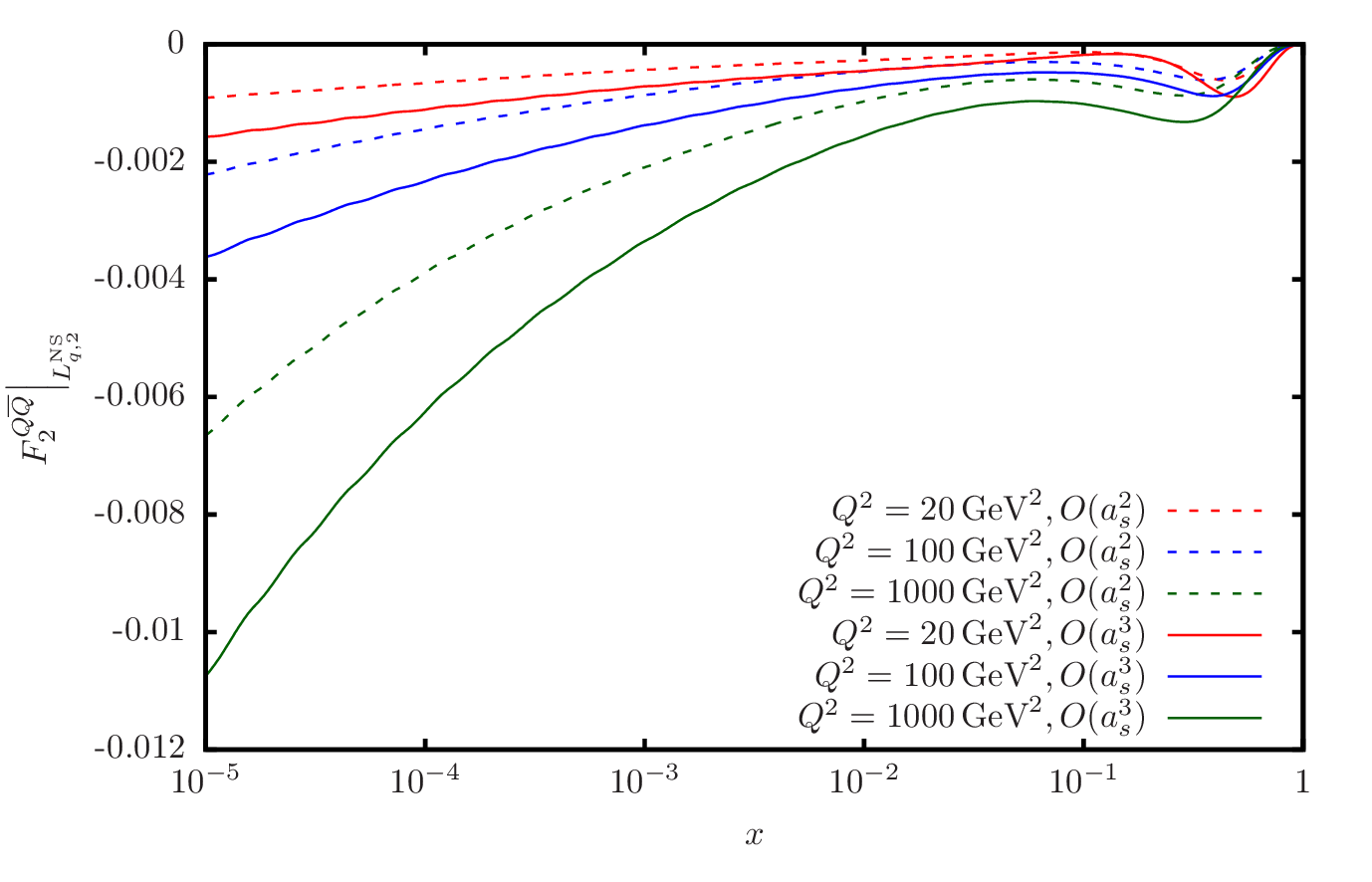}}
%{\epsfig{figure=mpsi2mc.eps,height=70mm}}
\caption{\scriptsize The flavor non-singlet contribution of the Wilson coefficient $L^{\rm NS}_{q,2}$ to the structure function
$F_2(x,Q^2)$ at 2- and 3-loop order using the ABM NNLO parton distribution functions (PDFs) \cite{Alekhin:2013nda}
in the on-shell scheme for $m_c=1.59$ GeV; from \cite{nonsinglet}.}
\label{NSgraph1} 
\end{figure} 
%---------------------------------------------------------------------------------------------------------------------------------------
%---------------------------------------------------------------------------------------------------------------------------------------
\begin{figure}[H]
\centering
\includegraphics[width=7cm]{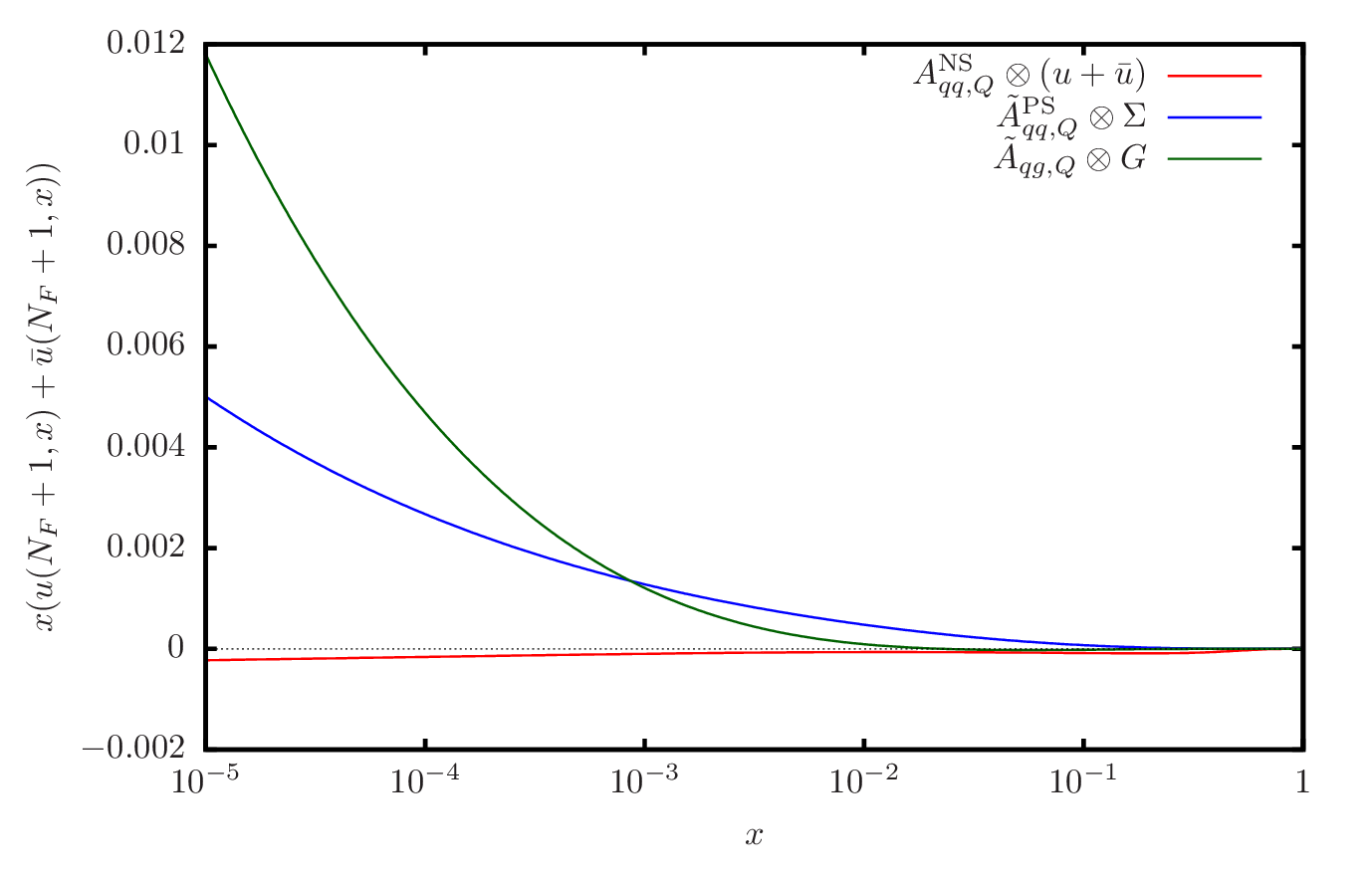}
\caption{\scriptsize The contributions to the distribution $x(u+\bar{u})$ at 3-loop order for four flavors in the variable 
flavor number scheme 
matched at the scale $\mu^2 = 20~{\rm GeV}^2$ using the 
on-mass-shell definition of the charm quark mass $m_c = 1.59~{\rm GeV}$ and using the PDFs \cite{Alekhin:2013nda}. The contributions due to 
the non-singlet, singlet and gluon distributions are shown individually; from \cite{nonsinglet}.}
\label{Fig:VFNS1}
\end{figure}
%---------------------------------------------------------------------------------------------------------------------------------------

The heavy flavor contributions to the structure function $F_2(x,Q^2)$ are obtained by a Mellin convolution of the heavy flavor Wilson 
coefficients with the respective parton distributions.
%---------------------------------------------------------------------------------------------------------------------------------------
\begin{figure}[hbt]
\centering
\includegraphics[width=7cm]{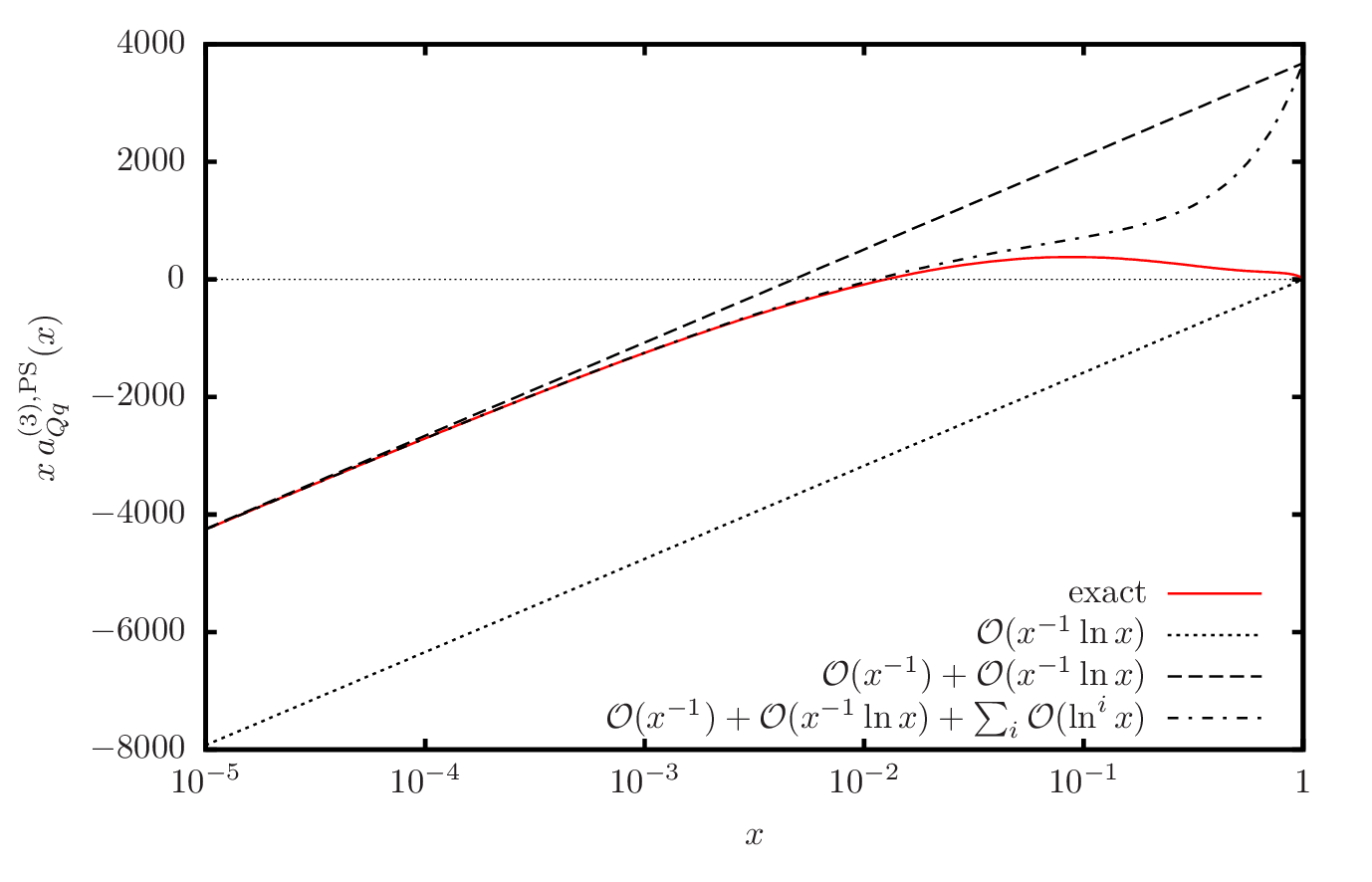}
\caption{\scriptsize $xa_{Qq}^{(3),\rm PS}(x)$ (solid red line) and leading terms approximating this 
quantity; dotted line: `leading' small $x$ approximation $O(\ln(x)/x)$, dashed line: adding the $O(1/x)$-term, dash-dotted
line: adding all other logarithmic contributions; from \cite{PSpaper}.}
\label{fig:1}
\end{figure}
%---------------------------------------------------------------------------------------------------------------------------------------
In Figure \ref{NSgraph1}, we show the contribution up to three loops of the non-singlet heavy flavor Wilson coefficient $L_{q,(2)}^{\rm NS}$ to $F_2(x,Q^2)$.
We can see that they are smaller than $1\%$ in the kinematic region of HERA. They will, however, become relevant at high luminosity 
machines such as the EIC, where
all 3-loop Wilson coefficients will be of importance.
As we mentioned in the introduction, the calculation of massive OMEs performed so far also allow us to obtain the transition relation for 
$n_f \rightarrow n_f+1$ 
massless flavors of the flavor non-singlet distribution in the VNFS. In Figure \ref{Fig:VFNS1}, we show the different contributions to the 4-flavor
distribution $x(u(x,\mu^2)+\bar{u}(x,\mu^2))$ as a function of $x$ for $\mu^2 = 20~{\rm GeV}^2$. We can see that the non-singlet 
contributions are very small, while
the singlet and gluon contributions are much larger for small values of $x$. The distributions for down and strange quarks are therefore 
nearly the same. 
%---------------------------------------------------------------------------------------------------------------------------------------
%---------------------------------------------------------------------------------------------------------------------------------------
\begin{figure}[H]
\centering
\includegraphics[width=6cm]{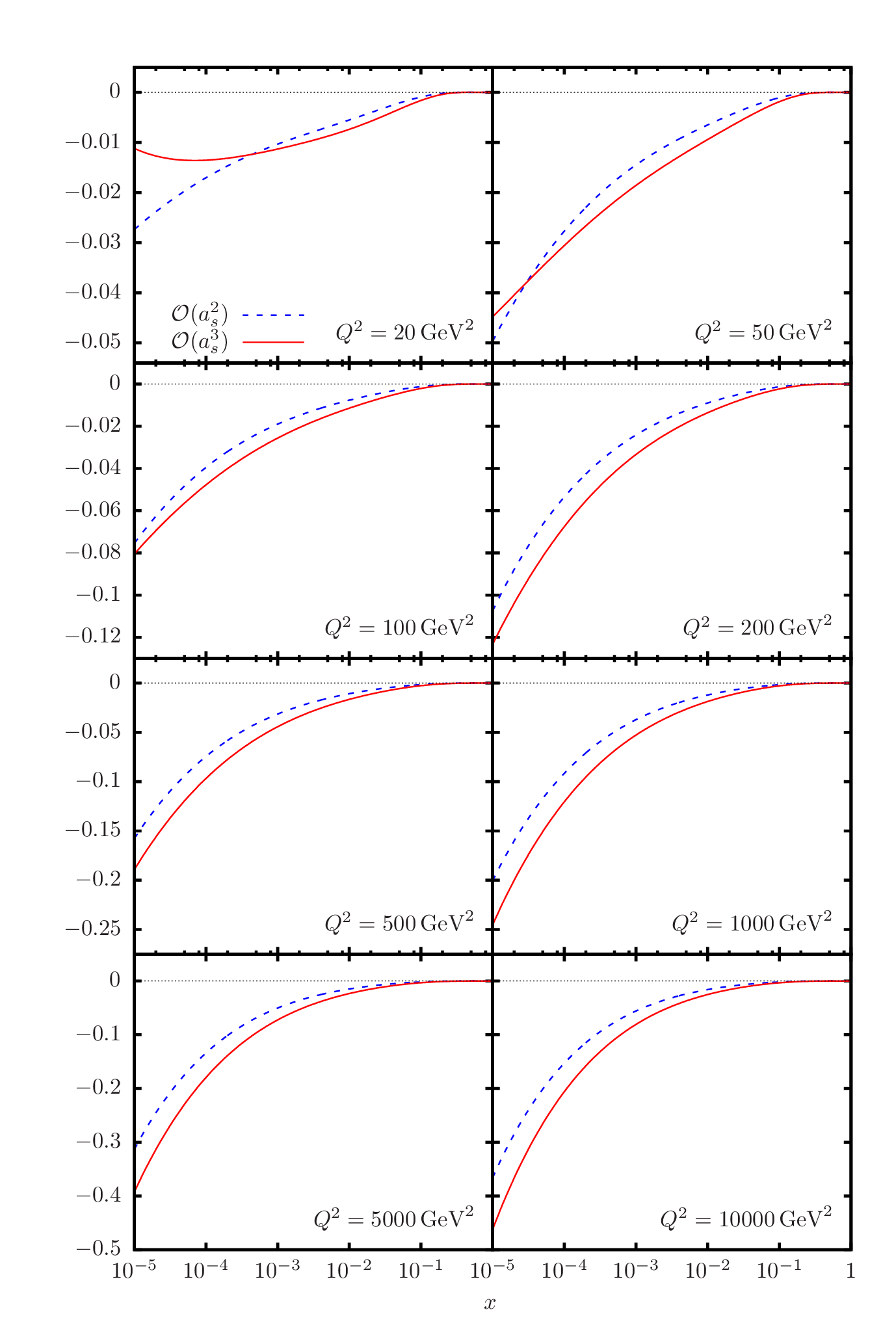}
\caption{\scriptsize The charm contribution by the Wilson coefficient $H_{q,2}^{\rm PS}$ to the structure function 
$F_2(x,Q^2)$ as a function of $x$ and $Q^2$ choosing $Q^2 = \mu^2, m_c = 1.59~{\rm GeV}$ (on-shell scheme) using the PDFs 
\cite{Alekhin:2013nda}; from \cite{PSpaper}.}
\label{FIG:F2c}
\end{figure}
%---------------------------------------------------------------------------------------------------------------------------------------

In Figure \ref{fig:1}, we show the behavior of the constant part of the unrenormalized pure singlet OME, $a_{Qq}^{(3), \rm PS}(x)$. The 
leading small $x$ term $\propto \ln(x)/x$ does nowhere describe this quantity. In the small $x$ region one has to add several subleading 
terms
to get a sufficient approximation. In Figure \ref{FIG:F2c}, the charm quark contribution to $F_2(x,Q^2)$ by the Wilson coefficient 
$H_{q,2}^{\rm PS}$ is shown at $O(a_s^2)$ and up to $O(a_s^3)$ in dependence on $x$ and several values of $Q^2$, setting the factorization 
scale 
$\mu^2=Q^2$. The $O(a_s^3)$ correction is quite significant in the region of low $x$, 
and at lower scales $Q^2$ the $O(a_s^2)$ corrections are larger than those at NNLO.

%---------------------------------------------------------------------------------------------------------------------------------------
\section{Conclusions}
\label{conclusions}
%---------------------------------------------------------------------------------------------------------------------------------------

\noindent
We have made considerable progress in recent years in the calculation of heavy flavor corrections to DIS at NNLO in
in the region of large virtualities. So far, we have calculated six out of eight massive operator matrix elements developing and applying 
a variety of mathematical and computer algebraic techniques. The results have been given in terms of (generalized) nested harmonic sums. 
As a by-product, we have calculated the terms proportional to $T_F$ in the 3-loop anomalous dimensions confirming results in the literature. 
The corresponding massive Wilson coefficients have been calculated and their contributions to $F_2(x,Q^2)$ determined and analyzed.

%---------------------------------------------------------------------------------------------------------------------------------------

%------------------------------------------------------------------------------------------------------------------------

\begin{thebibliography}{99}

%% \bibitem must have the following form:
%%   \bibitem{key}...
%%
%------------------------------------------------------------------------------------------------------------------------
%
%[1]
\bibitem{Bethke:2011tr}
  S.~Bethke et al., %, A.~H.~Hoang, S.~Kluth, J.~Schieck, I.~W.~Stewart, S.~Aoki, M.~Beneke and S.~Bethke {et al.},
  {\sf Proceedings of the 2011 Workshop on Precision Measurements of $\alpha_s$},
  arXiv:1110.0016 [hep-ph];\\
  %%CITATION = ARXIV:1110.0016;%%
%\bibitem{Moch:2014tta}
  S.~Moch, S.~Weinzierl et al., %S.~Alekhin, J.~Blumlein, L.~de la Cruz, S.~Dittmaier, M.~Dowling and J.~Erler {et al.},
  %``High precision fundamental constants at the TeV scale,''
  arXiv:1405.4781 [hep-ph].
%%CITATION = ARXIV:1405.4781;%%  
%-------------------------------------------------------------------------------------------------------
%
%[2]
\bibitem{Alekhin:2012vu}
  S.~Alekhin, J.~Bl\"umlein, K.~Daum, K.~Lipka and S.~Moch,
  %``Precise charm-quark mass from deep-inelastic scattering,''
  Phys.\ Lett.\ B {\bf 720} (2013) 172
  [arXiv:1212.2355 [hep-ph]].
  %%CITATION = ARXIV:1212.2355;%%
%------------------------------------------------------------------------------------------------------------------------
%
%[3]
\bibitem{EIC} 
C. Aidala et al. {\sf A High Luminosity, High Energy Electron-Ion-Collider}, 
A White Paper Prepared for the NSAC LRP 2007;
%\bibitem{Boer:2011fh}
  D.~Boer et al.
%, M.~Diehl, R.~Milner, R.~Venugopalan, W.~Vogelsang, D.~Kaplan, H.~Montgomery and S.~Vigdor {et al.},
  %``Gluons and the quark sea at high energies: Distributions, polarization, tomography,''
  arXiv:1108.1713 [nucl-th].
  %%CITATION = ARXIV:1108.1713;%%
%------------------------------------------------------------------------------------------------------------------------
%
%[4]
\bibitem{LHeC}
%\bibitem{AbelleiraFernandez:2012cc}
  J.~L.~Abelleira Fernandez {et al.}  [LHeC Study Group Collaboration],
  %``A Large Hadron Electron Collider at CERN: Report on the Physics and Design Concepts for Machine and Detector,''
  J.\ Phys.\ G {\bf 39} (2012) 075001
  [arXiv:1206.2913].
  %%CITATION = ARXIV:1206.2913;%%
%------------------------------------------------------------------------------------------------------------------------
%
%[5]
\bibitem{hera}
%\bibitem{Alekhin:2005dx}
  S.~Alekhin et al. %, G.~Altarelli, N.~Amapane, J.~Andersen, V.~Andreev, M.~Arneodo, V.~Avati and J.~Baines {et al.},
  %``HERA and the LHC: A Workshop on the implications of HERA for LHC physics: Proceedings Part A,''
  [hep-ph/0601012];
  %%CITATION = HEP-PH/0601012;%%
%\bibitem{Alekhin:2005dy}
%  S.~Alekhin, G.~Altarelli, N.~Amapane, J.~Andersen, V.~Andreev, M.~Arneodo, V.~Avati and J.~Baines {et al.},
  %``HERA and the LHC: A Workshop on the implications of HERA for LHC physics. Proceedings, Part B,''
  [hep-ph/0601013];
  %%CITATION = HEP-PH/0601013;%%
%\bibitem{Jung:2009eq}
  Z.~J.~Ajaltouni at al. %, S.~Albino, G.~Altarelli, F.~Ambroglini, J.~Anderson, G.~Antchev, M.~Arneodo and P.~Aspell {et al.},
  %``Proceedings of the workshop: HERA and the LHC workshop series on the implications of HERA for LHC physics,''
  [arXiv:0903.3861];\\
  %%CITATION = ARXIV:0903.3861;%%
%\bibitem{Dittmar:2005ed}
  M.~Dittmar et al. %, S.~Forte, A.~Glazov, S.~Moch, S.~Alekhin, G.~Altarelli, J.~R.~Andersen and R.~D.~Ball {\it et al.},
  %``Working Group I: Parton distributions: Summary report for the HERA LHC Workshop Proceedings,''
  hep-ph/0511119.
  %%CITATION = HEP-PH/0511119;%%
%------------------------------------------------------------------------------------------------------------------------
%
%[6]
\bibitem{Buza:1995ie}
  M.~Buza, Y.~Matiounine, J.~Smith, R.~Migneron and W.~L.~van Neerven,
  %``Heavy quark coefficient functions at asymptotic values Q**2 >> m**2,''
  Nucl.\ Phys.\ B {\bf 472} (1996) 611
  [hep-ph/9601302].
  %%CITATION = HEP-PH/9601302;%%
%------------------------------------------------------------------------------------------------------------------------
%
%[7]
\bibitem{MVV2005}
%\bibitem{Vermaseren:2005qc}
  J.A.M.~Vermaseren, A.~Vogt and S.~Moch,
  %``The Third-order QCD corrections to deep-inelastic scattering by photon exchange,''
  Nucl.\ Phys.\ B {\bf 724} (2005) 3
  [hep-ph/0504242].
  %%CITATION = HEP-PH/0504242;%%
%------------------------------------------------------------------------------------------------------------------------
%
%[8]
\bibitem{OurPapers1}
%\bibitem{Ablinger:2010ty}
  J.~Ablinger, J.~Bl\"umlein, S.~Klein, C.~Schneider and F.~Wi\ss{}brock,
  %``The O(\alpha_s^3) Massive Operator Matrix Elements of O(n_f) for the Structure Function F_2(x,Q^2) and Transversity,''
  Nucl.\ Phys.\ B {\bf 844} (2011) 26
  [arXiv:1008.3347 [hep-ph]];
  %%CITATION = ARXIV:1008.3347;%%
%\bibitem{Blumlein:2012vq}
  J.~Bl\"umlein, A.~Hasselhuhn, S.~Klein and C.~Schneider,
  %``The $O(\alpha_s^3 n_f T_F^2 C_{A,F})$} Contributions to the Gluonic Massive Operator Matrix Elements,''
  Nucl.\ Phys.\ B {\bf 866} (2013) 196
  [arXiv:1205.4184 [hep-ph]];
  %%CITATION = ARXIV:1205.4184;%%
%\bibitem{Blumlein:2014fqa}
  J.~Bl\"umlein, A.~Hasselhuhn and T.~Pfoh,
  %``The $O(\alpha_s^2)$ heavy quark corrections to charged current deep-inelastic scattering at large virtualities,''
  Nucl.\ Phys.\ B {\bf 881} (2014) 1
  [arXiv:1401.4352]; 
  %%CITATION = ARXIV:1401.4352;%%
%\bibitem{Ablinger:2014lka}
  J.~Ablinger, J.~Bl\"umlein, A.~De Freitas, A.~Hasselhuhn, A.~von Manteuffel, M.~Round, C.~Schneider and F.~Wi\ss{}brock,
  %``The Transition Matrix Element $A_{gq}(N)$ of the Variable Flavor Number Scheme at $O(\alpha_s^3)$,''
  Nucl.\ Phys.\ B {\bf 882} (2014) 263
  [arXiv:1402.0359].
  %%CITATION = ARXIV:1402.0359;%%
%------------------------------------------------------------------------------------------------------------------------
%
%[9]
\bibitem{Behring:2014eya}
  A.~Behring et al.
%, I.~Bierenbaum, J.~Bl\"umlein, A.~De Freitas, S.~Klein and F.~Wi\ss{}brock,
%  %``The Logarithmic Contributions to the O(\alpha_s^3) Asymptotic Massive Wilson Coefficients and Operator Matrix Elements in Deeply Inelastic Scattering,''
  [arXiv:1403.6356], Eur. Phys. Journ. C (2014) in print.
  %%CITATION = ARXIV:1403.6356;%%
%------------------------------------------------------------------------------------------------------------------------
%
%[10]
\bibitem{TF2paper}
%\bibitem{Ablinger:2014uka}
  J.~Ablinger, J.~Bl\"umlein, A.~De Freitas, A.~Hasselhuhn, A.~von Manteuffel, M.~Round and C.~Schneider,
  %``The O(\alpha_s^3 T_F^2) Contributions to the Gluonic Operator Matrix Element,''
  Nucl.\ Phys.\ B {\bf 885} (2014) 280
  [arXiv:1405.4259 [hep-ph]].
  %%CITATION = ARXIV:1405.4259;%%
%------------------------------------------------------------------------------------------------------------------------
%
%[11]
\bibitem{hyperlogs}
%\bibitem{Ablinger:2014yaa}
  J.~Ablinger, J.~Bl\"umlein, C.~Raab, C.~Schneider and F.~Wi\ss{}brock,
  %``Calculating Massive 3-loop Graphs for Operator Matrix Elements by the Method of Hyperlogarithms,''
  Nucl.\ Phys.\ B {\bf 885} (2014) 409
  [arXiv:1403.1137 [hep-ph]].
  %%CITATION = ARXIV:1403.1137;%%
%------------------------------------------------------------------------------------------------------------------------
%
%[12]
\bibitem{nonsinglet}
%\bibitem{Ablinger:2014vwa}
  J.~Ablinger et al.
%, A.~Behring, J.~Bl\"umlein, A.~De Freitas, A.~Hasselhuhn, A.~von Manteuffel, M.~Round, C.~Schneider, and F. Wi\ss{}brock,
  %``The 3-Loop Non-Singlet Heavy Flavor Contributions and Anomalous Dimensions for the Structure Function $F_2(x,Q^2)$ and Transversity,''
  Nucl.\ Phys.\ B {\bf 886} (2014) 733
  [arXiv:1406.4654 [hep-ph]].
  %%CITATION = ARXIV:1406.4654;%%
%------------------------------------------------------------------------------------------------------------------------
%
%[13]
\bibitem{PSpaper}
%\bibitem{Ablinger:2014nga}
  J.~Ablinger, A.~Behring, J.~Bl\"umlein, A.~De Freitas, A.~von Manteuffel and C.~Schneider,
  %``The 3-Loop Pure Singlet Heavy Flavor Contributions to the Structure Function $F_2(x,Q^2)$ and the Anomalous Dimension,''
  arXiv:1409.1135.
  %%CITATION = ARXIV:1409.1135;%%
%------------------------------------------------------------------------------------------------------------------------
%
%[14]
\bibitem{Bierenbaum:2009mv}
  I.~Bierenbaum, J.~Bl\"umlein and S.~Klein,
  %``Mellin Moments of the O(alpha**3(s)) Heavy Flavor Contributions to unpolarized Deep-Inelastic Scattering at Q**2 >> m**2 and Anomalous Dimensions,''
  Nucl.\ Phys.\ B {\bf 820} (2009) 417
  [arXiv:0904.3563]; 
%\bibitem{Blumlein:2009rg} 
  J.~Bl\"umlein, S.~Klein and B.~T\"odtli,
  %``O(alpha(s)**2) and O(alpha(s)**3) Heavy Flavor Contributions to Transversity at Q**2 >>m**2,''
  Phys.\ Rev.\ D {\bf 80} (2009) 094010
  [arXiv:0909.1547].
%------------------------------------------------------------------------------------------------------------------------
%
%[15]
\bibitem{Blumlein:2006mh}
  J.~Bl\"umlein, A.~De Freitas, W.~L.~van Neerven and S.~Klein,
  %``The Longitudinal Heavy Quark Structure Function F**Q anti-Q(L) in the Region Q**2 >> m**2 at O(alpha**3(s)),''
  Nucl.\ Phys.\ B {\bf 755} (2006) 272
  [hep-ph/0608024].
  %%CITATION = HEP-PH/0608024;%%
%------------------------------------------------------------------------------------------------------------------------
%
%[16]
\bibitem{Nogueira:1991ex}
  P.~Nogueira,
  %``Automatic Feynman graph generation,''
  J.\ Comput.\ Phys.\  {\bf 105} (1993) 279.
  %%CITATION = JCTPA,105,279;%%
%------------------------------------------------------------------------------------------------------------------------
%
%[17]
\bibitem{Tentyukov:2007mu}
  M.~Tentyukov and J.~A.~M.~Vermaseren,
  %``The Multithreaded version of FORM,''
  Comput.\ Phys.\ Commun.\  {\bf 181} (2010) 1419
  [hep-ph/0702279];
  %%CITATION = HEP-PH/0702279;%%
%\bibitem{Vermaseren:2000nd}
  J.~A.~M.~Vermaseren,
  %``New features of FORM,''
  arXiv:math-ph/0010025.
  %%CITATION = MATH-PH/0010025;%%
%------------------------------------------------------------------------------------------------------------------------
%
%[18]
\bibitem{reduze}
%\bibitem{vonManteuffel:2012np}
  A.~von Manteuffel and C.~Studerus,
  %``Reduze 2 - Distributed Feynman Integral Reduction,''
  [arXiv:1201.4330]; 
  %%CITATION = ARXIV:1201.4330;%%
%\bibitem{Studerus:2009ye}
  C.~Studerus,
  %``Reduze-Feynman Integral Reduction in C++,''
  Comput.\ Phys.\ Commun.\  {\bf 181} (2010) 1293
  [arXiv:0912.2546].
  %%CITATION = ARXIV:0912.2546;%%
%------------------------------------------------------------------------------------------------------------------------
%
%[19]
\bibitem{laporta}
%\bibitem{Laporta:1996mq}
  S.~Laporta and E.~Remiddi,
  %``The Analytical value of the electron (g-2) at order alpha**3 in QED,''
  Phys.\ Lett.\ B {\bf 379} (1996) 283
  [hep-ph/9602417].
  %%CITATION = HEP-PH/9602417;%%
%------------------------------------------------------------------------------------------------------------------------
%
%[20]
\bibitem{Ablinger:2012qm}
  J.~Ablinger, J.~Bl\"umlein, A.~Hasselhuhn, S.~Klein, C.~Schneider and F.~Wi\ss{}brock,
  %``Massive 3-loop Ladder Diagrams for Quarkonic Local Operator Matrix Elements,''
  Nucl.\ Phys.\ B {\bf 864} (2012) 52
  [arXiv:1206.2252 [hep-ph]].
  %%CITATION = ARXIV:1206.2252;%%
%------------------------------------------------------------------------------------------------------------------------
%
%[21]
\bibitem{SLATER}
L.J. Slater, {\sf Generalized Hypergeometric Functions}, (Cambridge University
Press, Cambridge, 1966).
%------------------------------------------------------------------------------------------------------------------------
%
%[22]
\bibitem{MELB}
E.W. Barnes, Proc. Lond. Math. Soc. (2) {\bf 6} (1908) 141; Quart.
Journ. Math. {\bf 41} (1910) 136; %–140
H. Mellin,
%"Abriß einer einheitlichen Theorie der Gamma- und der hypergeometrischen Funktionen."
Math. Ann. {\bf 68} (1910) 305;
%\bibitem{Smirnov2006}
V.~A. Smirnov, {\sf {Feynman Integral Calculus}},
(Springer, Berlin, 2006).
%------------------------------------------------------------------------------------------------------------------------
%
%[23]
\bibitem{Brown:2008um}
  F.~Brown,
  %``The Massless higher-loop two-point function,''
  Commun.\ Math.\ Phys.\  {\bf 287} (2009) 925
  [arXiv:0804.1660];
  %%CITATION = ARXIV:0804.1660;%%
%\bibitem{Panzer:2014caa}
  E.~Panzer,
  %``Algorithms for the symbolic integration of hyperlogarithms with applications to Feynman integrals,''
  arXiv:1403.3385 [hep-th].
  %%CITATION = ARXIV:1403.3385;%%
%------------------------------------------------------------------------------------------------------------------------
%
%[24]
\bibitem{DEQ}
%\bibitem{Kotikov:1990kg}
  A.~V.~Kotikov,
  %``Differential equations method: New technique for massive Feynman diagrams calculation,''
  Phys.\ Lett.\ B {\bf 254} (1991) 158;
  %%CITATION = PHLTA,B254,158;%%
%\bibitem{Caffo:1998yd}
  M.~Caffo, H.~Czyz, S.~Laporta and E.~Remiddi,
  %``Master equations for master amplitudes,''
  Acta Phys.\ Polon.\ B {\bf 29} (1998) 2627
  [hep-th/9807119];
  %%CITATION = HEP-TH/9807119;%%
%\bibitem{Caffo:1998du}
%  M.~Caffo, H.~Czyz, S.~Laporta and E.~Remiddi,
  %``The Master differential equations for the two loop sunrise selfmass amplitudes,''
  Nuovo Cim.\ A {\bf 111} (1998) 365
  [hep-th/9805118].
  %%CITATION = HEP-TH/9805118;%%
%------------------------------------------------------------------------------------------------------------------------
%
%[25]
\bibitem{Karr:81}
M.~Karr, {J.~ACM\/} {\bf 28} (1981) 305; %--350 
%\bibitem{Schneider:01}
C.~Schneider,  
%{\sf Symbolic Summation in Difference Fields\/} 
Ph.D. Thesis
RISC, JKU Linz technical report 01-17 (2001).
%------------------------------------------------------------------------------------------------------------------------
%
%[26]
\bibitem{SIG1}
C.~Schneider, {S\'em.~Lothar. Combin.\/} {\bf 56} (2007) 1,%--36 
 article B56b; 
%C.~Schneider, 
in:~{{Computer Algebra in Quantum Field Theory: Integration,
  Summation and Special Functions}\/} Texts and Monographs in Symbolic
  Computation eds. C.~Schneider and J.~Bl\"umlein  (Springer, Wien, 2013) 325 %--360
  [arXiv:1304.4134].
%------------------------------------------------------------------------------------------------------------------------
%
%[27]
\bibitem{harmonicsums}
%\bibitem{Ablinger:2013hcp}
  J.~Ablinger,
%  {\sf Computer Algebra Algorithms for Special Functions in Particle Physics}, Ph.D. Thesis, J. Kepler University Linz, 2012,
  [arXiv:1305.0687];
  %%CITATION = ARXIV:1305.0687;%%
%\bibitem{Ablinger:2010kw}
%  J.~Ablinger,
%  {\sf A Computer Algebra Toolbox for Harmonic Sums Related to Particle Physics}, Diploma Thesis, J. Kepler University Linz, 2009,
  [arXiv:1011.1176];
  %%CITATION = ARXIV:1011.1176;%%
%\bibitem{Ablinger:2011te}
  J.~Ablinger, J.~Bl\"umlein and C.~Schneider,
  %``Harmonic Sums and Polylogarithms Generated by Cyclotomic Polynomials,''
  J.\ Math.\ Phys.\  {\bf 52} (2011) 102301
  [arXiv:1105.6063].
  %%CITATION = ARXIV:1105.6063;%%
%------------------------------------------------------------------------------------------------------------------------
%
%[28]
\bibitem{Ablinger:2013cf}   
  J.~Ablinger, J.~Bl\"umlein and C.~Schneider,
  %``Analytic and Algorithmic Aspects of Generalized Harmonic Sums and Polylogarithms,''
  J.\ Math.\ Phys.\  {\bf 54} (2013) 082301
  [arXiv:1302.0378 [math-ph]].
  %%CITATION = ARXIV:1302.0378;%%
%------------------------------------------------------------------------------------------------------------------------
%
%[29]
\bibitem{EMSSP}
%\bibitem{Ablinger:2010pb}
  J.~Ablinger, J.~Bl\"umlein, S.~Klein and C.~Schneider,
  %``Modern Summation Methods and the Computation of 2- and 3-loop Feynman Diagrams,''
  Nucl.\ Phys.\ Proc.\ Suppl.\  {\bf 205-206} (2010) 110
  [arXiv:1006.4797];
  %%CITATION = ARXIV:1006.4797;%%
%\bibitem{Blumlein:2012hg}
  J.~Bl\"umlein, A.~Hasselhuhn and C.~Schneider,
  %``Evaluation of Multi-Sums for Large Scale Problems,''
  PoS (RADCOR 2011) 032
  [arXiv:1202.4303];
  %%CITATION = ARXIV:1202.4303;%%
%\bibitem{Schneider:2013zna}
  C.~Schneider,
  %``Modern Summation Methods for Loop Integrals in Quantum Field Theory: The Packages Sigma, EvaluateMultiSums and SumProduction,''
  J.\ Phys.\ Conf.\ Ser.\  {\bf 523} (2014) 012037
  [arXiv:1310.0160].
  %%CITATION = ARXIV:1310.0160;%%
%------------------------------------------------------------------------------------------------------------------------
%
%[30]
\bibitem{ORESYS}
S.~Gerhold, {\sf Uncoupling systems of linear {O}re operator equations},
Master's thesis, RISC, J.~Kepler University, Linz, 2002.
%------------------------------------------------------------------------------------------------------------------------
%
%[31]
\bibitem{HSUM}
%\bibitem{Blumlein:1998if}
  J.~Bl\"umlein and S.~Kurth,
  %``Harmonic sums and Mellin transforms up to two-loop order,''
  Phys.\ Rev.\  D {\bf 60} (1999) 014018
  [arXiv:hep-ph/9810241];\\
  %%CITATION = PHRVA,D60,014018;%%
%-----------------------------------------------------------------------------------
%\bibitem{Vermaseren:1998uu}
  J.A.M.~Vermaseren,
  %``Harmonic sums, Mellin transforms and integrals,''
  Int.\ J.\ Mod.\ Phys.\  A {\bf 14} (1999) 2037
  [arXiv:hep-ph/9806280].
  %%CITATION = IMPAE,A14,2037;%%
%------------------------------------------------------------------------------------------------------------------------
%
%[32]
\bibitem{Moch:2001zr}
  S.~Moch, P.~Uwer and S.~Weinzierl,
  %``Nested sums, expansion of transcendental functions and multiscale multiloop integrals,''
  J.\ Math.\ Phys.\  {\bf 43} (2002) 3363
  [hep-ph/0110083].
  %%CITATION = HEP-PH/0110083;%%
%------------------------------------------------------------------------------------------------------------------------
%
%[33]
\bibitem{Ablinger:2014bra}
  J.~Ablinger, J.~Bl\"umlein, C.~G.~Raab and C.~Schneider,
  %``Iterated Binomial Sums and their Associated Iterated Integrals,''
  arXiv:1407.1822 [hep-th].
  %%CITATION = ARXIV:1407.1822;%%
%------------------------------------------------------------------------------------------------------------------------
%
%[34]
\bibitem{Alekhin:2013nda}
  S.~Alekhin, J.~Bl\"umlein and S.~Moch,
  %``The ABM parton distributions tuned to LHC data,''
  Phys.\ Rev.\ D {\bf 89} (2014) 054028
  [arXiv:1310.3059 [hep-ph]].
  %%CITATION = ARXIV:1310.3059;%%
%------------------------------------------------------------------------------------------------------------------------
\end{thebibliography}
\end{document}